# An Empirical Analysis of Scam Tokens on Ethereum Blockchain

Vahidin Jeleskovic[1]

**Abstract**

This article presents an empirical investigation into the determinants of total revenue generated by counterfeit tokens on Uniswap. It offers a detailed overview of the counterfeit token fraud process, along with a systematic summary of characteristics associated with such fraudulent activities observed in Uniswap. The study primarily examines the relationship between revenue from counterfeit token scams and their defining characteristics, and analyzes the influence of market economic factors such as return on market capitalization and price return on Ethereum. Key findings include a significant increase in overall transactions of counterfeit tokens on their first day of fraud, and a rise in upfront fraud costs leading to corresponding increases in revenue. Furthermore, a negative correlation is identified between the total revenue of counterfeit tokens and the volatility of Ethereum market capitalization return, while price return volatility on Ethereum is found to have a positive impact on counterfeit token revenue, albeit requiring further investigation for a comprehensive understanding. Additionally, the number of subscribers for the real token correlates positively with the realized volume of scam tokens, indicating that a larger community following the legitimate token may inadvertently contribute to the visibility and success of counterfeit tokens. Conversely, the number of Telegram subscribers exhibits a negative impact on the realized volume of scam tokens, suggesting that a higher level of scrutiny or awareness within Telegram communities may act as a deterrent to fraudulent activities. Finally, the timing of when the scam token is introduced on the Ethereum blockchain may have a negative impact on its success. Notably, the cumulative amount scammed by only 42 counterfeit tokens amounted to almost 11214 Ether.

Key wards: Ethereum Blockchain, Fraud ICO, Scam Tokens, Empirical Analysis, Unisawp.

[1] Humboldt Universität zu Berlin, Spandauer Str. 1, D-10178 Berlin. Email: vahidin.jeleskovic@hu-berlin.de

## 1. Introduction

Scams continue to proliferate within the cryptocurrency market, with recent data indicating that crypto users lost approximately $4.2 billion in 2022 and around $2 billion in 2023.[2] These scams manifest in various forms, including fraudulent schemes, hacks, and exploits. Perpetrators demonstrate a high level of creativity and sophistication, adept at siphoning funds from unsuspecting users and investors in the crypto space.

Given these facts, establishing a standardized taxonomy of scams in the cryptocurrency realm poses a significant challenge. Conducting a systematic and comprehensive empirical analysis of the factors influencing the "success" of scams requires focusing on specific forms due to the sheer diversity and evolving nature of scams. Efforts in this direction are ongoing, driven by the vast array of scam types and cases, as well as the continual emergence of new forms.

A rough classification of mainstream cryptocurrency scams can be divided into three main types: hacking, fraudulent websites and bogus apps, and ICO scams. Hacking scams manifest themselves in a variety of ways, from attacking exchanges, mining companies or digital wallets, to installing malware on investors' computers to steal crypto wallets, such as WeSteal, a notable crypto-stealing malware. Hackers are further able to steal the sim information of mobile phones from investors in order to manipulate other people's cryptocurrency exchange accounts. Fraudulent websites and fake apps lure investors with promises of high returns or easy cryptocurrency mining, often disseminating false news to attract victims.

ICO scams, a newer form of fraud, exploit investors by creating a sense of exclusivity and potential profits. These scams typically involve promoting tokens through social media with misleading whitepapers or no blockchain backing, as seen in cases like Onecoin. Rug Pull scams, prevalent on decentralized exchanges (DEXs), exploit the decentralized nature of these platforms. Unlike centralized exchanges (CEXs), DEXs enable direct token trading among users, making them popular but also susceptible to scams. The simplicity and lack of regulation on DEXs render them vulnerable to token offering scams. Uniswap, currently a prominent DEX, boasted $47.37 billion in trading volume for Uniswap v1 and $24.07 billion for Uniswap v2 in November 2021. However, behind Uniswap's success lie numerous reported scams, particularly cases of fraud involving worthless token listings. This paper focuses on counterfeit tokens, where scammers mimic the names of upcoming crypto tokens with growth potential and list their fake tokens on Uniswap. They propagate false news via platforms like Telegram to entice investors into purchasing these counterfeit tokens, creating a temporary price spike to enhance credibility and deceive investors. Various schemes are devised to deceive investors into believing they are investing in genuine, newly listed crypto tokens.

The remainder of the paper is organized as follows. Section 2 conducts a literature review on fraudulent activities in the cryptocurrency market. Section 3 provides a detailed description of how counterfeit token scammers operate their schemes and explains the logic behind Uniswap's Automated Market Maker to understand the scammers' behavior thoroughly. Section 4 focuses on data sources, variable selection, descriptive statistics, and empirical results. The concluding section 5 summarizes the main findings and offers insights into the causes of these fraudulent activities.

## 2. Literature reviews

Research on cryptocurrency fraud is multifaceted. Vasek et al. (2015) categorize cryptocurrency scams into four main types: Ponzi schemes, mining scams, scam wallets, and fraudulent exchanges, based on their analysis of 192 bitcoin-related frauds. Chen et al. (2018) and Bartoletti et al. (2020) identify similar characteristics of Ponzi schemes in the context of Ethereum. According to AARP, common cryptocurrency frauds include celebrity endorsements, fake websites, virtual Ponzi schemes, and scams where imposters pose as legitimate traders or create fraudulent exchanges to swindle money from unsuspecting investors. Twomey et al. (2020) shed light on why cryptocurrency markets are vulnerable to manipulation, citing factors such as inconsistent regulation, relative anonymity, and low barriers to entry. Additionally, they highlight the lack of stringent procedures during exchange setups as contributing to this susceptibility.

---

[2] https://www.coindesk.com/tech/2023/12/27/crypto-users-lost-2b-to-hacks-scams-and-exploits-in-2023-defi-says/

Xia et al. (2021) categorize virtual cryptocurrency fraud into three main groups: security vulnerabilities in centralized exchanges, price manipulation in DeFi projects, and blockchain scams. They advocate for greater accountability in the cryptocurrency space, citing examples of bitcoin hacks and the closure of 18 crypto token exchanges between 2011 and 2017. Ji et al. (2020) identify vulnerabilities in ERC-20 smart contracts using an automated tool called 'depositafe', applicable to both centralized and decentralized exchanges. Kim and Lee (2018) analyze potential weaknesses in cryptocurrency exchanges and user wallets, addressing security policies but acknowledging risks associated with websites. McCorry et al. (2018) propose a reactive tool for hacked exchanges, similar to Bitcoin's vault, allowing for recovery key restoration and token destruction to mitigate losses. Feder et al. (2017) examine trading volume variations post-hacking of the MT.Gox exchange, noting reduced high-volume trading afterward. Xia et al. (2020) study fraudulent attacks on exchanges, identifying numerous fraudulent domains and applications resulting in significant financial losses. They focus on the security of centralized exchanges on preventing hacking attacks and fake dapps and apply an intensive descriptive analyzes figuring out corresponding factors and relationships within fraud scheme. In this line is also the publication by Chohan (2018). Corbet et al. (2020) find continued strong price fluctuations in major cryptocurrencies post-hack, indicative of reduced investor confidence. The aforementioned studies primarily focus on enhancing security measures for centralized exchanges and individual wallets post-hack, as well as assessing market impacts. Future research may explore scams on decentralized exchanges (DEXs), particularly ICO scams and fraudulent token issuance.

Shifflett et al. (2019) conducted a study analyzing the whitepapers of 3,300 crypto products online, identifying 513 of them as scam cryptocurrencies. Similarly, Zetzsche et al. (2017) found that over half of the 1,000 ICO projects in their database lacked essential information such as issuer details and token flow instructions, leading to skepticism about investing in these projects. Schueffel et al. (2019) assert that 2,2% of ICO cryptocurrencies are fraudulent based on their research, challenging Dowlat's (2018) claim that about 80% of ICO projects until July 2018 are scams. They propose the Crypto Scam Probability Index, focusing on factors like project funding, open-source code, developer activity, and KYC certification, for evaluating scam currencies. Tiwari et al. (2020) and Makarov and Schoar (2021) echo the need for market regulation. Xia et al. (2021) estimate that approximately 50% of tokens listed on Uniswap are scam tokens, resulting in losses of up to $16 million. Gao et al. (2020) found that most counterfeit tokens in Uniswap had minimal transaction activity, with 90% having fewer than 45 transactions. They identified 2,117 counterfeit tokens, 7,104 victims, and a total scam amount of $17 million. Mazorra et al. (2022) identified 27,588 rug pull scams in Uniswap, categorizing tokens based on smart contract features and liquidity pool status. Their analysis, based on data collected from Uniswap until September 2021, highlights the prevalence of fraudulent activities in decentralized exchanges.

The existing research on counterfeit tokens has strongly focused on algorithmic methods for detecting malicious addresses, with less attention given to understanding the underlying modus operandi of the fraud and the factors influencing the extent of counterfeit token scams. Given that counterfeit tokens traded on Uniswap are essentially valueless tokens transacted on the Ethereum blockchain, it's worth exploring whether fluctuations in the Ethereum market could affect the prevalence of counterfeit token scams. Studies have indicated a strong correlation between changes in the Ethereum market and trends in the broader cryptocurrency market. Nadler and Guo (2020) conducted an analysis of the cryptocurrency market using an asset pricing model, revealing that market-related risks, as well as risks associated with the Bitcoin and Ethereum blockchains, impact cryptocurrency prices. They found that the state of the Bitcoin blockchain has a significant influence on cryptocurrency pricing, particularly when the blockchain is not widely recognized. Additionally, as the number of ERC20 token-based cryptocurrencies grows, the Ethereum network effect gradually becomes more dominant. This suggests that developments in the Ethereum market could indeed play a role in influencing the dynamics of counterfeit token scams on platforms like Uniswap.

Sovbetov (2018) concludes that the volume-weighted indicator, derived from 50 selected crypto tokens representing 92% of the total cryptocurrency market capitalization, plays a significant role in determining short and long-term changes in Ethereum prices, showing a positive correlation. Additionally, the volatility-weighted indicator, based on the daily volatility of these 50 tokens, also exerts a notable influence on both Ethereum and Bitcoin price fluctuations across short and long-term periods. Angela and Sun (2020) gathered data on Ethereum and other altcoin prices in the cryptocurrency market from 2016 to 2018, employing an ARDL model to investigate the relationship between altcoin prices and Ethereum. Their analysis highlights the significant impact of three altcoins—Bitcoin, Litecoin, and Monero—on Ethereum prices. These findings offer valuable empirical evidence for identifying determinants of fraud volume. However, we will also offer additional insights and suggestions for this purpose.

Finally, Bartoletti et al. (2021) offer a comprehensive taxonomy for systematically categorizing various fraud and scam schemes related to cryptocurrencies and crypto tokens. They identify seven scam types: Ponzi schemes, Fake crypto services, Malware, Blackmail, Advance-fee scams, Money laundering, and Fake ICOs, delving deeply into each one. However, one type of scam scheme remains unexplored in empirical analysis: when scammers clone an ICO campaign for a legitimate token on the Ethereum blockchain and Uniswap, using all available channels, but swap the address of the legitimate token with that of their fake token. Hence, we analyze a specific type of scam which we refer to as 'Cloned ICO'. Doing this, we use the term ICO collectively for Initial Coin Offerings, Initial Exchange Offerings (IEO), and Initial DEX Offerings (IDO), as all these methods aim to raise funds by offering new tokens. Despite their operational differences—ranging from the platforms they utilize to the level of regulatory oversight and access for participants—the core objective of these strategies is to introduce and sell cryptocurrency tokens to support project development. This umbrella term simplifies the discussion around these varied fundraising mechanisms, focusing on their shared goal rather than the specifics of their execution.

## 3. Description of the automated market maker (AMM) protocols and the procedures of the counterfeit token scheme

**3.1 Description of the AMM protocols**

The decentralized exchange (DEX) not only facilitates trading for existing cryptocurrencies but also enables the creation of new ones directly within its market system. Its distributed structure shields the exchange from interference by local or international authorities. Additionally, the unique Automated Market Maker (AMM) protocols further enable fraudulent activities such as counterfeit tokens.[3]

AMM involves pooling everyone's assets into a liquidity pool and making markets based on an algorithm known as the Constant Product Market Maker Model (CPM). This model ensures that the product of the token pairs in the liquidity pool remains constant before and after a transaction, providing a mechanism for seamless trading on platforms like Uniswap.[4]

---

[3] https://docs.uniswap.org/protocol/V2/concepts/protocol-overview/glossary

[4] By "constant product", it means practically "XY = Z", where Z is always a constant value, no matter how X and Y change. In the context of Uniswap transactions, this means that the product of the two tokens in the liquidity pool before and after a transaction remains always constant, i.e., the pre-buy product = the post-buy product. For example: create a new liquidity pool of ETH and counterfeit tokens in Uniswap (free to create, no fees), the number of ETH and counterfeit tokens at creation is 10 and 20 respectively, the price of counterfeit tokens at this time is 0.1 ETH, the product of the two quantities is 10*20=200. to exchange ETH for the counterfeit token.

In Uniswap's AMM protocols, users can also add funds to the liquidity pool to help reduce the slippage of transactions. However, if tokens are added to the liquidity pool at random, the ratio of the number of tokens will change and the price will fluctuate significantly. In order to keep the current ratio of the two token prices from being changed, the user needs to add the current ratio of both tokens to the liquidity pool at the same time. In this way, the product is expanded, but the exchange price of the two tokens does not change.

**3.2 Procedures of the counterfeit token scheme**

The scheme involves four straightforward steps. Firstly, utilizing existing network resources like https://remix.ethereum.org, to construct the framework of the counterfeit token, enabling its query on the Ethereum blockchain. Secondly, establishing a liquidity pool on the Uniswap platform with the counterfeit token and a commonly used liquid fiat currency counterpart, such as USDT. Thirdly, injecting Ethereum's native currency, Ether, into the created pool to manipulate the price of the counterfeit token. Due to the nature of the Automated Market Maker (AMM) mechanism on Uniswap, token prices can fluctuate rapidly in response to investors' actions. Consequently, the price of counterfeit tokens can surge quickly due to the Ether injected by fraudsters, giving the impression that the counterfeit coin is a newly launched, promising token. This attracts more investors to purchase the counterfeit token promptly.

The fourth step involves disseminating false information on platforms like Telegram and Twitter. Scammers often create official accounts for counterfeit tokens to deceive investors into believing they are following legitimate token updates. Picture 1 in the appendix illustrates scammers posing as official token staff, sharing updates about counterfeit tokens on Telegram. These steps are typically executed simultaneously during fraudulent operations, leaving investors with irreparable losses once they realize they have purchased counterfeit tokens. It's noteworthy that fraudsters employ various tactics to enhance the credibility of counterfeit tokens. Firstly, they time the launch of counterfeit tokens around the ICO date of the genuine token they are imitating. Secondly, the counterfeit token's name is usually identical to the genuine token's, as per the classification by Gao et al. (2020). Additionally, the working address name of the counterfeit token closely resembles that of the genuine token. The names of the 42 counterfeit tokens analyzed in this study, along with the names of the genuine tokens they replicate, are provided in the appendix.[5]

**4. Research methodology**

**4.1 Data collection**

The research question of this paper revolves at first around investigating whether fluctuations in the price of Ethereum, which holds a significant share of the Uniswap market, impact the volume of counterfeiter tokens. The primary contribution lies in studying the specific characteristics of counterfeit token scams and the factors influencing the extent of these frauds. When a particular cryptocurrency garners enough investor interest, its market price typically experiences a surge. The absence of trading time restrictions in cryptocurrency markets implies that price fluctuations occur continuously. On the Uniswap platform, price fluctuations of crypto tokens present investment opportunities for traders. In the case of counterfeit tokens, fraudsters manipulate various aspects such as price fluctuations, token

---

Suppose someone uses 1 ETH to buy counterfeit tokens, at this point 2Eth enters the liquidity pool and the number of ETH becomes 12, then to keep the product of 200 constant, the number of counterfeit tokens has to be reduced by counterfeit tokens. This reduced number of counterfeit tokens is the number of counterfeit tokens that can be bought with 2 ETH. According to the constant product, 10*20=(10+2)*(20-counterfeit tokens'), the calculated counterfeit tokens' = 3.33, i.e. 1 ETH can buy 3.33 counterfeit tokens, and the slippage (price error) relative to the original price of 1 ETH=2counterfeit tokens is ( 2- 3.33)/2*100%  = 66.5%.

[5] The observation, we made in our dataset, that more than 70% of projects opted for an Initial Coin Offering (ICO) rather than an IEO or IDO highlights a significant trend in the cryptocurrency fundraising landscape.

issuance quantity, and transaction volume to deceive investors. The total value of counterfeit tokens throughout the scam cycle comprises the scam costs and investor funds. Notably, the fraudsters recycle their initial investment at the end of the scam by transferring all funds, including the initial input costs and fraudulent gains, back to their own wallet address. This process exploits Uniswap's Automated Market Maker mechanism, wherein fraudsters initially invest Ether to artificially inflate the price of counterfeit tokens, enhancing the credibility of the scam. However, once a predetermined profit threshold is met, fraudsters swiftly withdraw all funds from the counterfeit token's account to their personal wallet address.

Tokens with high overall rankings across various categories on cryptocurrency platforms are more susceptible to counterfeiting. Therefore, the manual data collection process focuses on identifying tokens with high overall rankings in different categories on prominent cryptocurrency information platforms. This study primarily utilizes the platform https://coinmarketcap.com to identify crypto tokens with high aggregate rankings in various categories. Once a list of tokens with high rankings is compiled, their names are manually entered into the search box of Dextools.app, a search engine based on Uniswap trading information. Leveraging Dextools.app's search algorithm, liquidity pools associated with the entered token names are retrieved. By excluding pools related to authentic token transactions, pools created by counterfeiters on Uniswap are identified.

This methodology identifies numerous counterfeit token addresses, although not all scams are successful. Only counterfeit tokens that are actively traded indicate successful implementation of the scam. Consequently, a final collection of 42 counterfeit token addresses was obtained, meeting specific criteria. These criteria include:

1. The number of transaction records exceeding 10.

2. The first transaction record of the counterfeit token not being significantly distant from the corresponding authentic token.

3. Preference for collecting counterfeit tokens with investor comments containing keywords such as "scammer," "fake address," "give my money back," etc., on the counterfeit token's address homepage. An example of a token commented on by investors as a fraud can be viewed in picture 2 of the appendix. This is an indicate that the fraud was "successful".

After collecting the addresses of 42 eligible counterfeiter tokens and analyzing their data, several characteristics of the counterfeiter token scam emerge. Firstly, the fraud cycle is brief, with the peak transfer period occurring on the first day the genuine token is listed. This finding aligns with previous research by Xia et al. (2021), Gao et al. (2020), and Vasek et al. (2018). Secondly, fraudsters promptly inject Ethereum into their tokens to manipulate prices once they are listed. Thirdly, counterfeit tokens are often de-addressed (or "burned") at the end of the scheme period,[6] making the scam challenging to trace. Fourthly, counterfeit tokens may become untradeable or unsellable after purchase. Fifthly, the average number of transactions for counterfeiter tokens is typically less than 45, consistent with findings by Gao et al. (2020). Lastly, scammers typically transfer the initial costs and fraudulently obtained amounts back to their wallet address at the end of the scamming process, consistent with research by Torres et al. (2019) on counterfeiters' characteristics.

**4.2 Regression model**

In our study, informed by a comprehensive literature review and our own analytical considerations, we have categorized key factors believed to influence the success of ICOs, which also shed light on the dynamics surrounding the volume of scam tokens. These factors are primarily liquidity, marketing and promotion strategies, the nuances of electronic trading, and some relevant features of the ecosystem which is in this caste the Ethereum blockchain. By establishing the log-values of the first-day realized volume of scam tokens as our dependent variable, labeled "volume_scam", we identify variables based on their anticipated causal impact on the dependent variable, prior to undertaking regression analysis.

---
[6] https://swissborg.com/blog/token-burning

In analyzing the liquidity factor's impact on scam token dynamics, we focus on "the cost of the scam," which we define as the quantity of Ethereum that fraudsters deposit into the liquidity pool to facilitate their fraudulent activity. This metric is denoted as "cost_scam," and we apply the logarithm to it for analysis. This approach allows us to quantify the initial investment fraudsters commit to creating the appearance of legitimate market activity for their scam tokens. The "cost_scam" variable serves as a crucial indicator of the resources fraudsters are willing to allocate to establish their scam, underpinning the liquidity they inject into the market to make the scam appear more credible to potential victims. It's premised on the understanding that a higher initial liquidity might attract more investors by portraying a semblance of market trust and token stability, and so to have positive effect on the realized volume of the scam token.

In the context of electronic trading microstructure, particularly with respect to high-frequency trading, empirical studies have shown a positive correlation between trading volume and both market volatility and trading intensity, the latter being quantified by the number of trades. This relationship is supported by research from Manganelli (2005), Hautsch (2007), and Hautsch and Jeleskovic (2008). Given our analysis focuses on a single day, squarely placing it within the realm of high-frequency trading, we hypothesize that the volume of scam tokens is similarly positively correlated with indicators of volatility and trading intensity. To assess volatility, we employ proxies based on the absolute price return of Ethereum (with price measured in dollars) on the days when fraudsters launch their tokens on Uniswap, denoted as "abs_ret_E". This choice reflects the hypothesis that fluctuations in Ethereum's price could influence scam token volume positively by affecting overall market sentiment and liquidity. To mitigate scaling effects and enhance the interpretability of our analysis, we adjust the "abs_ret_E" variable by multiplying it by 100.

To further refine our analysis of market volatility's impact on scam token dynamics, we introduce another volatility proxy: the absolute price returns of Ethereum relative to Bitcoin, with this measure also being scaled by 100 for consistency. This adjustment ensures comparability across different data points and facilitates a more nuanced understanding of market movements. We label this adjusted metric as "abs_ret_EB." The rationale behind using the absolute price returns of Ethereum in relation to Bitcoin is to capture significant market shifts that could influence investor behavior. A higher activity level in the Ethereum/Bitcoin pair may divert investor attention and capital away from alternative investments, including scam tokens. By scaling this proxy, we aim to provide a clear, quantifiable measure of this diversion effect. We hypothesize that an increase in Ethereum's volatility, as indicated by "abs_ret_EB," may have a deterrent effect on the volume of scam tokens. This is based on the premise that heightened volatility and trading interest in major cryptocurrency pairs like Ethereum/Bitcoin could lead investors to prioritize these more established assets over riskier, less-known tokens, thereby reducing the influx of investments into scam tokens. This adjusted variable allows us to test the assumption of a negative relationship between market volatility in major pairs and the appeal of scam tokens to potential investors.

For trading intensity, we use the log-values of the total number of trades within the first day of the fraud as our proxy, labeled "trans_numb". This metric aims to capture the immediacy and fervor of trading activity, under the assumption that higher trading volumes, spurred by intense trading activity, are positively indicative of a more significant presence of scam tokens.

To assess the impact of the Ethereum ecosystem's performance on new token dynamics, we employ a measure of the absolute change in Ethereum's market capitalization, expressed in Bitcoin. This metric serves as an indicator of Ethereum's ecosystem power and its competitive stance relative to Bitcoin, the most dominant cryptocurrency. Recognizing the complex relationship between Ethereum's market performance and its influence on the proliferation of new tokens—including scam tokens—we opt for a nonlinear approach to capture this dynamic more accurately. Accordingly, we transform the variable by taking the logarithm of the absolute change in market capitalization, which we denote as "log(abs_ret_MC)". This logarithmic transformation helps in smoothing out the effects of large fluctuations and provides a more nuanced view of how variations in Ethereum's market capitalization relative to Bitcoin may affect the environment for launching new tokens. The use of the logarithm underscores our understanding that the relationship might not be directly proportional and allows for a more refined analysis of how shifts in the Ethereum ecosystem's performance could influence the attractiveness and feasibility of introducing scam tokens to the market. This approach aims to offer deeper insights into the economic underpinnings that govern the crypto market's dynamics, particularly how Ethereum's standing impacts the broader token landscape.

To analyze the impact of marketing and promotional efforts on the dynamics of new token launches, including the potential for scam token proliferation, we consider the number of subscribers associated with the genuine token's promotional channels as a key metric. The underlying rationale is that a higher subscriber count reflects greater visibility and attractiveness of the token to potential investors, which inadvertently could increase the likelihood of individuals encountering and potentially falling prey to scam tokens associated with or mimicking the genuine token. This subscriber count serves as a proxy for the effectiveness of a token's marketing strategies and its ability to engage and expand its investor base. We denote this metric as "token_subskr." The assumption is that the broader the audience a real token can captivate through its marketing and promotional efforts, the larger the pool of potential victims for scam tokens. In further examining the influence of marketing and promotional efforts on the dynamics of real versus scam tokens, we consider the engagement metrics of genuine tokens, particularly focusing on their community engagement on social media platforms. A pivotal aspect of this analysis is the number of subscribers to the official channel of the real token on Telegram, a platform widely recognized for its role in promoting new ICOs and fostering community engagement. This metric serves as a proxy for the strength and cohesion of the real token's community, underpinning our hypothesis that a robust community may act as a deterrent against the allure of scam tokens which could, in turn, decrease the likelihood of community members falling prey to scam tokens. This variable is labeled as "telegram_subskr". Both variables are logarithmically transformed.

Timing is crucial for the success of this type of scam. If fraudsters release their scam token on the Ethereum blockchain before the legitimate token, they increase the likelihood of deceiving naive investors by substituting the wrong token address for the correct one in their promotional campaigns. However, if this incorrect address remains in circulation for an extended period, they also raise the risk of being exposed as scammers. Therefore, they face a trade-off. We measure the timing by the variable 'ico_lag,' which represents how long before the official ICO the fraudulent token was introduced on the Ethereum blockchain. We apply the logarithm to it for analysis. We expect a negative impact on the volume of the scam token, as investors may become more cautious due to the early exposure of the fraudulent activity.

**4.3 Empirical results**

Table 1 provides descriptive statistics for the realized volume of scam tokens ("valume_scam") and its associated features while Table 2 presents the regression results.

The data reveals that the average total value of counterfeit tokens on the first day of the scam stands at approximately 235 ETH, with a minimal average total value of counterfeit tokens in circulation throughout the entire scam period, amounting to merely 267 ETH. This suggests that a significant portion of transactions occurs within the initial day of the entire fraudulent cycle.

**Table 1: Counterfeit token descriptive statistics**

|  | valume_scam on first day | valume_scam | number of trasactions on first day | number of trasactions | the cost of the scam | scam interval |
|---|---|---|---|---|---|---|
| Mean | 234.97 | 266.99 | 75.83 | 89.928 | 168.72 | 18.31 |
| Median | 215.46 | 237.65 | 33.50 | 43.50 | 124.03 | 5 |
| Maximum | 911.88 | 1030.17 | 462 | 474.00 | 1012 | 333 |
| Minimum | 1.75 | 1.75 | 4 | 4 | 0.160 | 1 |
| Std. Dev. | 210.61 | 242.26 | 111.55 | 117.07 | 185.11 | 51.21 |
| Skewness | 1.49 | 1.46 | 2.50 | 2.23 | 2.44 | 5.71 |
| Kurtosis | 5.42 | 5.10 | 8.42 | 6.88 | 11.38 | 35.61 |
| Jarque-Bera | 25.81 | 22.75 | 95.31 | 61.21 | 164.75 | 2090.65 |
| Probability | 2e-2 | 1.1e-5 | 0.000 | 0.000 | 0.000 | 0.000 |
| Sum | 9868.80 | 11213.96 | 3185.00 | 3777.00 | 7086.36 | 769 |
| Obs. | 42 | 42 | 42 | 42 | 42 | 42 |

**Table 2: Empirical results**

| Variable | Coefficient | Std. Error | t-Statistic | Prob. |
|---|---|---|---|---|
| C | 4.322673 | 1.259806 | 3.431220 | 0.0018 |
| cost_scam | 0.419656 | 0.049662 | 8.450188 | 0.0000 |
| abs_ret_EB | -0.656826 | 0.207734 | -3.16186 | 0.0036 |
| abs_ret_E | 0.122160 | 0.060458 | 2.020585 | 0.0251 |
| log(abs_ret_MC) | 0.531442 | 0.185518 | 2.864630 | 0.0076 |
| trans_numb | 0.362937 | 0.080858 | 4.488567 | 0.0001 |
| token_subskr | 0.038005 | 0.015417 | 2.465046 | 0.0196 |
| telegram_subskr | -0.000102 | 4.34E-05 | -2.36114 | 0.0249 |
| ico_lag | -0.407354 | 0.169180 | -2.40781 | 0.0224 |
| $R^2$ | 0.867317 | | | |
| Adjusted $R^2$ | 0.831935 | | | |
| F-statistic | 2.451284 | Akaike info criterion | | 1.565756 |
| Prob(F-statistic) | 0.000000 | Schwarz criterion | | 1.949655 |

Equally noteworthy is the cost incurred by scammers in perpetrating counterfeit token schemes, with an average swindling cost for the counterfeit tokens amounting to 168.72 ETH. Similarly, the majority of transactions occurred predominantly on the first day. Analyzing the duration of the scam intervals, we note a wide range, spanning from a minimum of one day to a maximum of 333 days before the official start of the ICO. This variability suggests diverse strategies among fraudsters. Some may have aimed to secure a token name similar to the legitimate one by posting their scam token early, while others could be attributed to delays in the ICO of the authentic token.

From Table 3, several key conclusions emerge. Firstly, there's no evidence of heteroskedasticity in the residuals across any of the models, indicating robustness in the model's performance. Secondly, all coefficient estimates are statistically different from zero, signifying their significance. Specifically, a 1% increase in the number of transactions of counterfeit tokens on their first official fraud day corresponds to an estimated rise of approximately 0.36% in the total fraud amount on that day. Likewise, a 1% increase in the fraudulent cost of counterfeit tokens on their first official fraud day leads to an estimated increase of approximately 0.41% in the total fraud amount on the first scam day.

Moreover, it's noteworthy that the absolute return of Ethereum in dollars exhibits a positive impact, while the absolute return of Ethereum in Bitcoin demonstrates a negative impact, aligning with expectations. Additionally, the statistically significant coefficient signs for token and Telegram subscribers align with expectations. Finally, ICO lag exerts a notably large negative effect.

## 5. Conclusion

In the previous section, it was established that upfront costs and transaction volume of counterfeit tokens positively influence their total revenue. A higher transaction volume indicates greater liquidity in the counterfeit token pool during the scam period. For scammers, who act as liquidity providers, this liquidity ensures they receive proceeds from the project. They must maintain the counterfeit token price within a certain range to ensure revenue. Therefore, scammers strive to maximize pool liquidity during the scam, adjusting token weights and providing additional Ether to the pool as needed. Increased costs and transactions make it more profitable for scammers to invest more to maintain higher returns and attract investor funds.

Furthermore, the success of scam token projects is contingent upon market and ecosystem conditions. This article examines counterfeit tokens in the cryptocurrency trading market from a fraudulent perspective, revealing several key characteristics. Firstly, the trading cycle is short-term, with peak activity occurring 1 to 2 days after real tokens are listed on Uniswap. Counterfeit tokens yield high returns at minimal cost, apart from the initial investment required to create a liquidity pool between the counterfeit token and a crypto stablecoin. However, ultimately, all invested capital plus fraudulently obtained funds flow into the pockets of the scammers. Therefore, from the perspective of the scammers, higher initial capital investment in the liquidity pool and increased swapping of counterfeit tokens for Ether lead to higher returns.

Overall, several factors contribute to the proliferation of counterfeit tokens regarding the Uniswap exchange, and this article proposes the following explanations. Firstly, the absence of costs associated with fraud significantly lowers the barriers for scammers. Uniswap allows any user to set up a liquidity pool for free, eliminating token listing fees and reducing the overall cost of perpetrating fraud. Consequently, scammers face minimal obstacles, whether setting up pairing liquidity pools, executing medium-term fraudulent activities, or abandoning counterfeit tokens altogether.

Secondly, the Uniswap Automated Market Maker (AMM) mechanism exacerbates the issue. Given the relatively small liquidity pools maintained by many counterfeiters, trading exchange rates fluctuate rapidly. A sharp increase in price can occur within a short period, attracting a surge of investors. When potential investors observe such price spikes in counterfeit tokens, they may be misled into believing they are accessing the correct token address.

However, a strong community can mitigate the impact of scams, underscoring the importance of thorough due diligence by investors. Additionally, regulators are urged to establish better infrastructure and standardized systems for Initial Coin Offerings (ICOs). There is also a call for stricter identification requirements for individuals behind new projects, enhancing transparency and accountability in the cryptocurrency space.

**Appendix**

Table 1: *The name comparison between corresponding real token and counterfeit token*

| Real Token Name | Scam token name | Working address name |
|---|---|---|
| aleph.im | aleph.im | Uniswap V2: ALEPH 4 |
| Antimatter.Finance Governance Token | Antimatter.Finance Governance Token | Uniswap V2: MATTER 5 |
| APYSwap | APYSwap | Uniswap V2: APYS |
| Bankroll Vault | Bankroll Vault | Uniswap V2: BVLT 11 |
| BarnBridge Governance Token | BarnBridge Governance Token | Uniswap V2: BOND 12 |
| CARD.STARTER | CARD.STARTER | Uniswap V2: CARDS 2 |
| Clover | Clover | Uniswap V2: LUCK 5 |
| Coin98 | Coin98 | Uniswap V2: C98 3 |
| Disbalancer | Disbalancer | Uniswap V2: DDOS 2 |
| DVGToken | DAOVentures | Uniswap V2: DVG 8 |
| Elongate Deluxe | Elongate Deluxe | |
| Fei USD | Fei Protocol [IDO] | Uniswap V2: TRIBE 6 |
| Tribe | Fei Tribe | Uniswap V2: TRlBE |
| Fractal Protocol Token | Fractal Protocol.id | Uniswap V2: FCL 3 |
| Frax Share | Frax Share | Uniswap V2: FXS 4 |
| Furucombo | Furucombo | Uniswap V2: COMBO 6 |
| Gitcoin | Gitcoin | Uniswap V2: GTC 3 |
| HORD Token | HORD | Uniswap V2: HORD 2 |
| Instadapp | Instadapp | Uniswap V2: INST |
| InsurAce | InsurAce | Uniswap V2: INSUR |
| keep4r | keep4r | Uniswap V2: KP4R 3 |
| Lattice Token | Lattice Token | Uniswap V2: LTX 12 |
| Launchpool token | LaunchPool | Uniswap V2: LPOOL 2 |
| Litentry | Litentry | Uniswap V2: LIT 9 |
| Meter Governance mapped by Meter.io | Meter Governance mapped by Meter.io | Uniswap V2: eMTRG |
| Modefi | Modefi | Uniswap V2: MOD 4 |
| O3 Swap Token | O3 Swap Token | Uniswap V2: O3 3 |
| Plasma | Plasma | Uniswap V2: PPAY 27 |
| Public Mint | Public Mint | Uniswap V2: MINT 7 |
| Quick Swap | QuickSwap | Uniswap V2: QUICK |
| Sekuritance | Sekuritance | Uniswap V2: SKRT |
| SuperFarm | SuperFarm | Uniswap V2: SUPER 10 |
| Tidal Token | Tidal Token | Uniswap V2: TlDAL |
| Unido | Unido | Uniswap V2: UDO |
| UniLend Finance Token | UniLend Finance | Uniswap V2: UFT 14 |
| UnmarshalToken | UnmarshalToken | Uniswap V2: MARSH 8 |
| VAIOT Token | VAIOT | Uniswap V2: VAI 6 |
| Yearn Finance Diamond Token | Yearn Finance Diamond token | Uniswap V2: YFDT 4 |
| Power Index Pool Token | PowerIndex | Uniswap V2: PIPT 2 |

Picture 1: An example of a counterfeit HOLOTOKEN (HOT) airdrop scam found on social media. These posts closely mimic legitimate token airdrop announcements, but with the distinction of featuring a fraudulent token address.

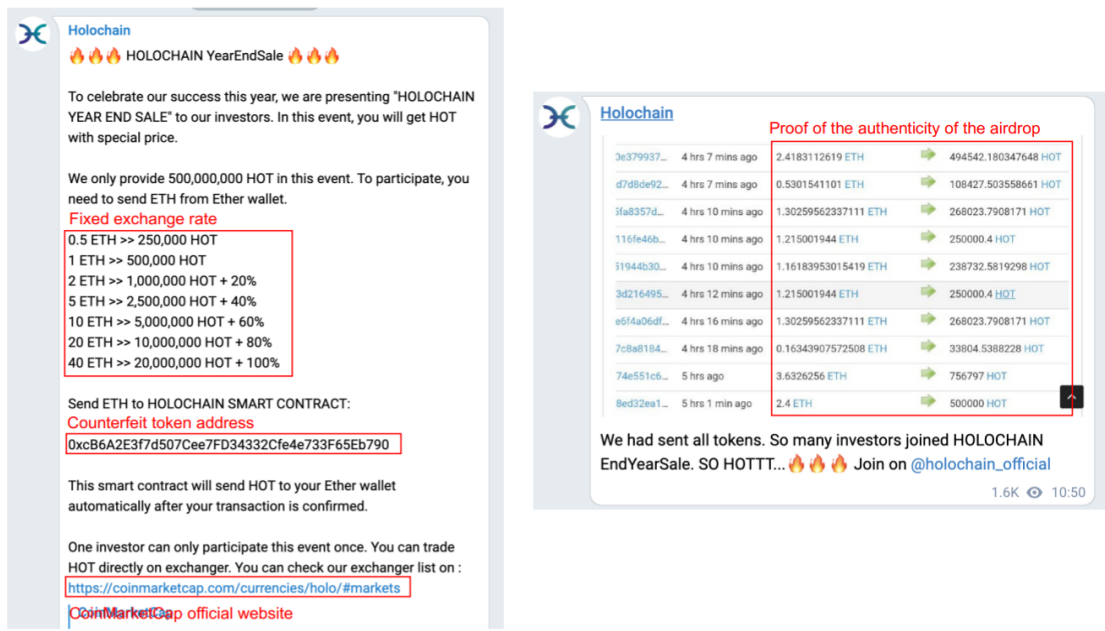

Picture 2: *Comments from the defrauded investors*
*(Source: https://etherscan.io/token/0x7b1385e62febf72dc17e2c5804605d476ad404ff#comments)*

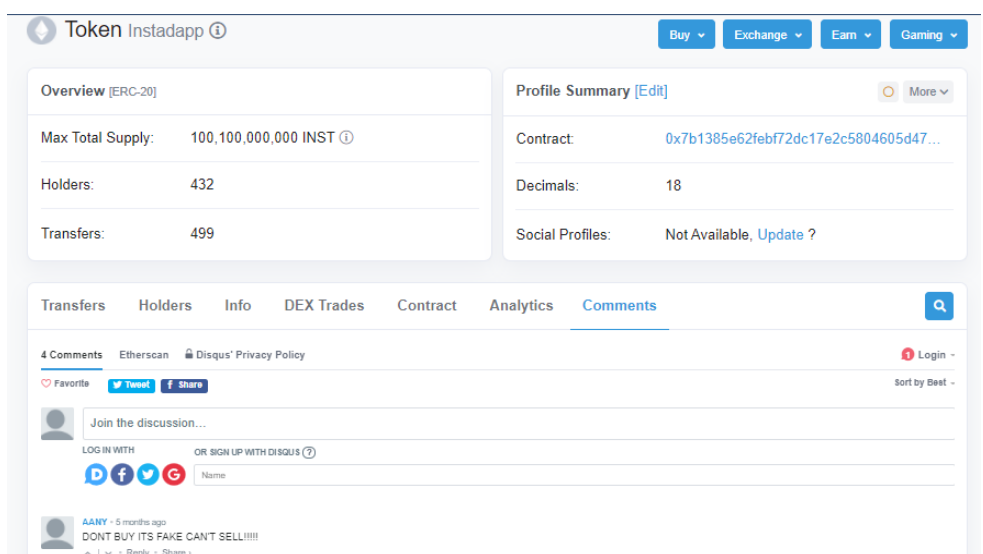

Picture 3: *Dextools.app search box autofill feature on display*

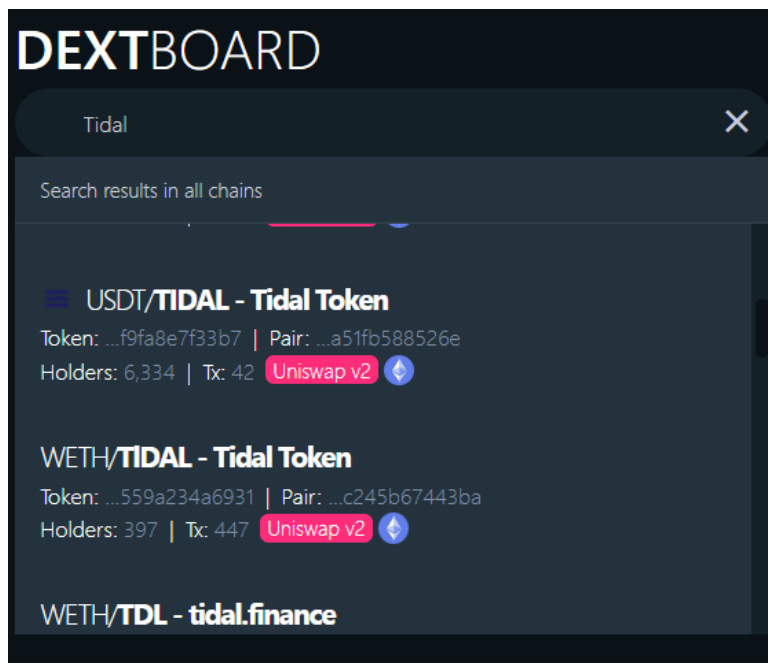